**Quantum Limits in Nanomechanical Systems**

In two articles [1, 2], the authors claim that the Heisenberg uncertainty principle limits the precision of simultaneous measurements of the position and velocity of a particle and refer to experimental evidence that supports their claim.

It is true that ever since the inception of quantum mechanics, the uncertainty relation that corresponds to a pair of observables represented by non-commuting operators is interpreted by many scientists and engineers, including Heisenberg himself, as a limitation on the accuracy with which observables can be measured [3-5]. However, such a limitation cannot be deduced from the postulates and theorems of quantum thermodynamics.

The measurement result theorem avers that the measurement of an observable is a precise (perturbation free) and, in many cases, precisely calculable eigenvalue of the operator that represents the observable. So neither a measurement perturbation nor a measurement error is contemplated by the theorem. An outstanding example of measurement accuracy is the Lamb shift correction of energy eigenvalues [6]. The probability theorem avers that we cannot predict which precise eigenvalue each measurement will yield except in terms of either a prespecified or measurable probability or frequency of occurrence.

It follows that an ensemble of measurements of an observable performed on an ensemble of identical systems, identically prepared yields a range of eigenvalues, and a probability or frequency of occurrence distribution over the eigenvalues. In principle, both results are precise and involve no disturbances induced by the measuring procedures.

To be sure, each probability distribution of an observable represented by operator X has a variance

$$(\Delta X)^2 = Tr\rho X^2 - (Tr\rho X)^2 = \langle X^2 \rangle - \langle X \rangle^2$$

and a standard $\Delta X$, where $\rho$ is the projector or density operator that describes all the probability distributions of the problem in question. Moreover, for two observables represented by two non-commuting operators A and B, that is,

$$AB - BA = iC$$

it is readily shown that $\Delta A$ and $\Delta B$ satisfy the uncertainty relation [7-9]

$$\Delta A \Delta B \geq |\langle C \rangle|/2$$

It is evident that each uncertainty relation refers neither to any errors introduced by the measuring instruments nor to any particular value of a measurement result. The reason for the latter remarks is that the value of an observable is determined by the expectation value of the operator representing the observable and not by any individual measurement result. In addition, these results are valid for all probability catalogs, that is, projectors and density operators that can be represented by a homogeneous ensemble, including those that correspond to thermodynamic equilibrium at any temperature T $\left(-\infty \leq 1/T \leq \infty\right)$.

Elias P. Gyftopoulos
Ford Professor Emeritus, Departments of Mechanical
and Nuclear Engineering, Massachusetts Institute of Technology
77 Massachusetts Avenue, Cambridge, MA 02139, USA
E-mail: epgyft@aol.com